\documentclass[structabstract]{aa}

\usepackage{epsfig,graphicx,latexsym,amsmath,amssymb}
\usepackage{natbib}
\usepackage{hyperref}
\usepackage{mathrsfs}
\usepackage{lastpage}
\usepackage{color}

\bibpunct[,]{(}{)}{;}{a}{}{,}
\begin{document}

 \author{
Symeon Konstantinidis\inst{1,5}\thanks{e-mail: Simos@ari.uni-heidelberg.de},
Pau Amaro-Seoane\inst{2}\thanks{e-mail: Pau.Amaro-Seoane@aei.mpg.de}
\& Kostas D. Kokkotas\inst{3,4}\thanks{e-mail: Kostas.Kokkotas@uni-tuebingen.de}
}

\institute{Astronomisches Rechen-Institut, M{\"o}nchhofstra{\ss}e 12-14, 69120,
Zentrum f\"ur Astronomie, Universit\"at Heidelberg, Germany \and
Max Planck Institut f\"ur Gravitationsphysik
(Albert-Einstein-Institut), D-14476 Potsdam, Germany \and
Theoretical Astrophysics (TAT), IAAT, Eberhard Karls University of T{\"u}bingen,
Auf der Morgenstelle 10, 72076 T{\"u}bingen, Germany \and
Department of Physics, Aristotle
University of Thessaloniki, Thessaloniki 54124 Greece \and
Departamento de Astronom\'\i a y Astrof\'\i sica, Facultad de F\'{i}sica,
Pontificia Universidad Cat\'olica de Chile,
Av. Vicu\~{n}a Mackenna 4860, 782-0436 Macul, Santiago, Chile
}

\authorrunning{S. Konstantinidis, P. Amaro-Seoane and K. D. Kokkotas}
\titlerunning{Kicking IMBHs off GCs: IMRIs}

\date{\today}

\label{firstpage}

\title{
       Investigating the retention of intermediate-mass black holes in star clusters
       using $N$-body simulations
       }

\begin{abstract}
{Contrary to supermassive and stellar-mass black holes (SBHs), the existence of
intermediate-mass black holes (IMBHs) with masses ranging between
$10^{2-5}\,M_{\odot}$ has not yet been confirmed.  The main problem in the
detection is that the innermost stellar kinematics of globular clusters (GCs) or small galaxies,
the possible natural loci to IMBHs, are very difficult to resolve.  However, if IMBHs
reside in the centre of GCs, a possibility is that they interact dynamically
with their environment. A binary formed with the IMBH and a compact object of
the GC would naturally lead to a prominent source of gravitational radiation,
detectable with future observatories.}
{We use $N$-body simulations to study the evolution of GCs containing an IMBH and
calculate the gravitational radiation emitted from dynamically formed IMBH-SBH binaries and
the possibility that the IMBH escapes the GC after an IMBH-SBH merger.}
{We run for the first time direct-summation integrations of GCs with an IMBH including the dynamical
evolution of the IMBH with the stellar system and relativistic effects, such as
energy loss in gravitational waves (GWs) and periapsis shift, and gravitational
recoil.}
{We find in one of our models an intermediate mass-ratio inspiral
(IMRI), which leads to a merger with a recoiling velocity higher than the
escape velocity of the GC. The GWs emitted fall in the range of frequencies
that a LISA-like observatory could detect, like the European eLISA or in
mission options considered in the recent preliminary mission study conducted in
China.  The merger has an impact on the global dynamics of the cluster, as an
important heating source is removed when the merged system leaves the GC.  The
detection of one IMRI would constitute a test of GR, as well as an irrefutable
proof of the existence of IMBHs.}
{}
\end{abstract}


\maketitle

 \section{Motivation}
 \label{sec.motivation}

 Intermediate-mass black holes (IMBHs), with masses $M\sim 10^{2-5}~ \rm M_\odot$
 possibly exist at the centres of globular clusters (GCs) or small galaxies,
 if we assume that they follow the observed correlations
 between super-massive BHs (SMBHs) and their stellar surroundings \citep[see][ and
 references therein]{TremaineEtAl02,GuelketinEtAl09,MillerColbert04,Miller09}.
 Due to their mass, these objects cannot form
 via the
 formation scenarios of stellar-mass black holes,
 which are the final results of stellar evolution of massive stars \citep{Fryer1999,FK01,BelczynskiEtAl2010}
 or SMBHs with masses $M\sim 10^{6-9}~\rm M_\odot$, that exist at the centres of galaxies
 \citep{Rees78,Rees84,GillessenEtAl09} and for which there exists an emerging consensus about
 their formation
 \citep{VolonteriRees05,MandauRees2001}.

 There have been proposed several scenarios for the formation of IMBHs
 most of which require extremely dense environments, similar to the centers of GCs,
 for the IMBHs to form and grow in mass \citep{vanderMarel2004, MillerColbert04}.
 \cite{MH02} suggest in their work that such massive black holes (BHs) can form from repeated
 mergers of a $\sim 50 \, \rm M_\odot$ BH, located at the center of
 a GC, with other SBH of lower mass. The $\sim 50 \, \rm M_\odot$ threshold
 is required to ensure that the BH will not receive large recoil
 velocities after each merger and so will remain bound to the GC.
 According to \citep{MH02} the initial $\sim 50 \, \rm M_\odot$ BH
 could be formed either directly from the collapse of a massive star, or from
 a large number of SBH-SBH mergers, which would produce mostly
 escaping SBHs, but also a minority of large BHs, bound to the GC.

 An interesting scenario for the formation of IMBHs in the early
 evolution of GCs, has been studied by \cite{QuinlanShapiro1990,PortegiesZwartMcMillan2002,
 PortegiesZwartMcMillan00,GurkanEtAl04,PortegiesZwartEtAl04,FreitagEtAl06}.
 According to this scenario, the most massive stars sink to the centre of a GC even before
 they become BHs and thus the cluster experiences an early core collapse
 during which the central density of stars becomes large enough, that massive stars start
 to rapidly and continuously merge with each other \citep[see also][]{PortegiesZwartEtAl04,GoswamiEtAl2012}.
 This runaway process very soon leads to the formation of a very massive star (VMS), located
 close to the centre of the GC. It is unknown how stellar
 evolution proceeds in such a VMS \citep{GlebbeekEtAl2009}, but if it is assumed that the star
 directly collapses to a BH, without significant mass-loss, this could form an IMBH. Accretion
 of stars and gas during the next Myrs could increase its mass up to two orders of magnitude
 \citep{VesperiniEtAl2010}. Finally, as in SMBHs, Population III stars have been proposed as
 possible progenitors of IMBHs \citep[see][ and references therein]{vanderMarel2004,WhalenFryer2012},
 but there are still many uncertainties in the evolution of such a star \citep{HegerWoosley2002}.

 Although the formation of IMBHs has been studied extensively during the
 last decades and their existence has been proposed in the early 70s
 \citep{Wyller1970,BahcallOstriker1975,FrankRees1976},
 there is still no direct proof of their existence. However, there is an increasing
 number of favouring evidences that suggest that they should exist. The most prominent evidence is
from the observations of ultra-luminous X-ray sources \citep[ULXs,][]{FengEtAl2011}, which are usually
associated to IMBHs. The brightest known ULX, known as HLX-1, is located in
the outskirts of the edge-on S0a type galaxy
ESO243-49 and is currently the strongest IMBH candidate. Based on
the extreme luminosity of the X-ray source, which has a maximum of up to $1.1\times10^{42}~{\rm erg\,s^{-1}}$
in the $0.2-10$ keV band,
\cite{FarrellEtAl2009}
derive a conservative lower limit of $500\,\rm M_{\odot}$ for the potential IMBH
\citep[see also][]{GodetEtAl2009,FarrellEtAl2010}.
More recent observations measured a peak luminosity of $1.3\times10^{42}~{\rm erg\,s^{-1}}$
 \citep{GodetEtAl2011a} and a possible period of variability of $\sim 1 \rm yr$
 \citep{GodetEtAl2012,ServillatEtAl2011}.
X-ray luminosities up to $\sim 10^{41}~{\rm erg\,s^{-1}}$ can be explained
by super-Eddington accretion to $\sim 20 \, \rm M_{\odot}$ SBHs \citep{Begelman2002} and/or
beaming \citep{King2008}. However, larger BH masses are needed for explaining luminosities
$ > 10^{41}~{\rm erg\,s^{-1}}$. For HLX-1 the most recent estimate for the mass of the potential
IMBH is $\sim 3\times 10^3 - 10^5 \, \rm M_\odot$ \citep{GodetEtAl2012,ServillatEtAl2011,DavisEtAl2011},
very well in the range of masses of IMBHs.
Further investigations of HLX-1 confirmed its extraordinary luminosity
by proving its association with galaxy ESO243-49 at a distance of 95 Mpc \citep{WiersemaEtAl2010},
and thus made the evidence of an IMBH even stronger. Interestingly, the X-ray source
is not located at the center of the host galaxy, but it lies at a distance
$ \sim 3.3 \, \rm kpc$ from its center and $\sim 0.8 \, \rm kpc$ out of the galactic plane,
possibly associated with a star cluster which appears to be in the same area.
According to \cite{FarrellEtAl2012} this cluster has a mass of $\sim 4 \times 10^6 \, \rm M_\odot $
and is either a massive young star cluster or an old GC.
Optical observations of HLX-1 with VLT seem to rule out the case of a massive star
cluster and favour the presence of a $\sim 10 \, \rm Gyr$ old globular cluster
with mass $< 10^6 \, \rm M_\odot$ or a $<10 \, \rm Myr$ small star cluster with mass
$\sim 10^4 \, \rm M_\odot$ \citep{SoriaEtAl2010,SoriaEtAl2012}.

Other recent interesting
observational examples that point to the existence of these objects can be found in the
work of \cite{SuttonEtAl2012}, which evaluates a sample of eight extreme
luminosity ultra-luminous X-ray source candidates and state that the observed
luminosities can be explained in terms of IMBHs with masses in the range of
$10^3-10^4\, \rm M_{\odot}$. Another X-ray source that might
be associated with an IMBH is found at the center of the
nearby (${\rm d} = 3.1 \, \rm Mpc$, \cite{KarachentsevEtAl2004}) dwarf lenticular galaxy
NGC 404 \citep{BinderEtAl2011}. Using both stellar and gas dynamical
mass estimates, \cite{SethEtAl2010} estimated the mass of the potential IMBH to be $\sim 10^5 \, \rm M_\odot$,
which agrees with recent estimates from Expanded Very Large Array
observations \citep{NylandEtAl2012} and from X-ray observations \citep{BinderEtAl2011}.
Finally, \cite{NylandEtAl2012} confirmed
the location of the source at the center of the nuclear star cluster hosted
by NGC 404 and ruled out other possible scenarios such as an X-ray binary,
stellar formation or a supernova remnant.

 The above observational examples provide strong evidence of the existence of IMBHs,
 but do not indisputably prove that they exist.
 A direct proof would come from detailed
 kinematical observations of stars moving under the influence of the
 IMBH at the centers of GCs. Unfortunately, the radius of influence of an IMBH is only of a
 few arc seconds \citep{Peebles72,ChanameEtAl10,MillerColbert04}, so
 it is very difficult, if not impossible, to accurately determine its mass
 by measuring the velocities of stars moving under its influence,
 with the currently available instruments.
 This technique has been successfully used for determining
 the mass of the SMBH at the centre of the Milky Way galaxy, where the stellar
 environment is less dense than the core of GCs and also there exists a number of
 young and bright stars, moving under the  gravitational influence of the SMBH which have
 been followed by observations for more than 15 years \citep{GillessenEtAl09}.
 The radius of influence $R$ of an IMBH of mass $M_\bullet$ can be defined as:
\begin{equation}
 R = \frac{G M_\bullet}{\sigma^2},
\end{equation}
where $\sigma$ the velocity dispersion at the center of the cluster. At a distance
$d$ this translates to an angular radius of influence \citep{Bender2005}:
\begin{equation}
\alpha = 1'' \Big( \frac{M_\bullet}{10^3 \ {\rm M_\odot}} \Big)
\Big( \frac{\sigma}{10 \ {\rm km\ s^{-1}}} \Big)^{-2} \Big( \frac{d}{10 \ {\rm kpc}} \Big)^{-1}
\end{equation}
 For a $10^4\,\rm M_{\odot}$ IMBH the influence radius is of $\sim 5''$, assuming a
 central velocity dispersion of $\sigma = 20 \ {\rm km\,s^{-1}}$ and a distance
 of $\sim 5$ kpc \citep[see also][ for a similar example]{MillerColbert04}. Also,
 since GCs are old systems, this small sphere of influence contains mainly massive
 stellar remnants and old, dim stars that could not be easily observed and traced.
 For the above reasons, kinematical techniques can currently only give upper limits on the mass
 of the potential IMBHs at the centers of galactic GCs
 \citep{AndersonvanderMarel2010,AndersonvanderMarel2010b,NoyolaEtAl2010,LutzgendorfEtAl2012}
 \citep[see also][ for observations that do not support the IMBH scenario]{KirstenVlemmings2012,StraderEtAl2012}.
 Since such limits are based on measurements of proper motions, velocity dispersion or
 line of sight motions away of the sphere of influence of the potential IMBH,
 alternative to IMBH explanations cannot be ruled out \citep{BHMMcMPZ03a,BaumgardtEtAl2005}.
 Hence, we would need the Very Large Telescope Interferometer (VLTI) and one of the
 next-generation near-infrared instruments, the VSI or GRAVITY
 \citep{GillessenEtAl06,EisenhauerEtAl08}.
 In that case we could improve the astrometric accuracy by an order of magnitude and thus we would
 possibly be in the position of detecting the innermost kinematics of a GC
 around a potential IMBH and thus measure accurately its mass.

An interesting avenue towards the {\em direct}
detection of an IMBH, which would not require future optical or infrared telescopes and
several years of observations, is GW astronomy.  Additionally, IMBH-SBH binaries that
might form in GCs represent an
excellent test of GR, since they are similar to extreme mass-ratio inspirals
\citep{Amaro-SeoaneEtAl07}. In particular, space-borne detectors such as
the ESA-led eLISA \citep{Amaro-SeoaneEtAl2012} or
Chinese mission study options \citep[``ALIA'' from now onwards,
see][]{BenderEtAl05,CrowderCornish05,GongEtAl11} will be able to catch these
systems (which might also be referred as intermediate mass-ratio inspirals, IMRIs)
with good signal-to-noise ratios (SNR) if the GC is not further than $z
\sim 0.7$ \citep{Amaro-SeoaneEtAl2012,MH02,Miller2006}. According to \cite{MH02},
LISA will be able to detect around 10 IMBH-SBH binaries at any given time, while
the merger of the BHs might be detectable by LIGO-II \citep[and Advanced LIGO should see
many of them][]{Amaro-SeoaneSantamaria10,FregeauEtAl06}.

 If an IMRI forms in a GC, it is undoubted that sooner or later
 it will lead to an IMBH-SBH merger. Recent studies from numerical relativity
 \citep{KoppitzEtAl2007,PollneyEtAl2007,RezzollaEtAl2008,Rezzolla09,LoustoEtAl10,LoustoZlochower11a,LoustoZlochower11b}
 show that BH-BH mergers result to a gravitational wave recoil
 which, depending on the mass-ratio and spins of the merging BHs, might
 be as large as $\sim 5000 \ \rm km \ s^{-1}$ \citep{LoustoZlochower11b}. The mass-ratio
 of an IMRI in a GC is large enough to avoid such large recoils, but it is still
 possible for an IMBH of mass up to $\sim 10^3 \ \rm M_\odot$ to receive a kick greater than the escape velocity
 of the GC and therefore leave the system \citep{Holley-Bockelmann08}.

 In this work we use $N$-body simulations to study the interactions of an IMBH
 with SBHs in young star clusters and describe, for the first time, the production on an IMRI
with our direct-summation code in one of our integrations. In Section \ref{sec:Tool} we describe the
 numerical tool and choice of the initial data used for the simulations. In
 Section \ref{sec:Dynamics} we describe the interactions of the IMBH with SBHs
 in the simulation we observed an IMRI, and we discuss the possibility that
 the gravitational recoil velocity assigned to the IMBH after the merger, kicks
 the IMBH out of the GC. In Section \ref{sec:GW} we calculate the gravitational
 radiation from such an IMRI in an approximate way. Finally, in Section \ref{sec:Conclusions}
 we conclude our work showing that an IMRI would be detectable by future
 space-based GW detectors, such as LISA, we discuss the
 effects of the ejection of the IMBH on the GC, their
 possible connection with ULXs not associated with GCs and we present
 our future plans for a statistical study of IMBH-SBH interactions in GCs.

 \section{Numerical tool and initial conditions}
 \label{sec:Tool}

We integrate the dynamical evolution of a globular cluster containing
a $500 - 1000 \ \rm M_\odot$ IMBH with \texttt{Myriad}
\citep{KonstantinidisKokkotas10}, a direct-summation $N-$body code that
integrates all gravitational forces for all particles at every time step. The
programme uses the Hermite integration scheme \citep{Aarseth99,Aarseth03}. This
requires computation of not only the accelerations, but also their time
derivatives. Particles that are tightly bound or with very small separation are
integrated using the time-symmetric Hermite scheme \citep{KokuboEtAl98}, which
is a symplectic integrator that makes the numerical errors oscillate between
two limits that can be controlled by the choice of the time step. The code uses
post-Newtonian correcting terms to the Newtonian forces, including 1, 2 and 2.5
order, as described for the first time in an $N-$body code by \cite{KupiEtAl06}
(their equations 1, 2 and 3), as well as a recipe for gravitational recoil.
The recoil velocity depends strongly on the mass ratio of the two holes, on the
magnitude of their spins and on their directions with respect to the plane of
the orbit \citep[see e.g.][ and references therein]{Rezzolla09}.  The equation
that we have implemented in the code is taken from \cite{LoustoEtAl10},

\begin{equation} \label{recoil}
 \vec{v} = \left(v_m + v_\bot \cos{\xi}\right)\, \hat{e}_1 + v_\bot \sin{\xi}\, \hat{e}_2 + v_{\parallel}\, \hat{e}_3.
\end{equation}

\noindent
In the last equation, the indices $\bot$ and $\parallel$ stand for
perpendicular and parallel directions with respect to the orbital angular
momentum vector $\vec{L}$ of the binary. $\hat{e}_1$ is a unit vector and lies
on the plane of the orbit connecting the two MBHs, with direction from the
heavier to the lighter one. $\hat{e}_2$ is also on the plane of the orbit, but
perpendicular to $\hat{e}_1$, with direction such that $\hat{e}_1$, $\hat{e}_2$
and $\hat{e}_3$ construct an orthonormal system, with $\hat{e}_3$ defined such
that it is the unit vector parallel to $\vec{L}$. $\xi$ is the angle  between the
unequal contributions of mass and spin to the recoil velocity.  We assign
random, maximal spins to the stars of the GC, and in particular we initially
give the IMBH a spin $a = S/M^2$  \citep[see e.g.][]{LoustoEtAl10} of 0.998.

We assume that the IMBH forms at the center of the cluster
when the GC is $10 \ \rm Myr$ old. This agrees with the formation scenario of
runaway stellar mergers \citep{PortegiesZwartMcMillan2002} and also of the
repeated SBH-SBH mergers \citep{MH02}. For our study the number and masses
of SBHs are of particular importance. Therefore, before creating the
initial data for our simulations, we studied the number of SBHs and their
masses assuming different initial mass functions (IMF) and metallicities.
{ For the initial mass function (IMF) we use Kroupa-like distributions \citep[see][]{KroupaEtAl93,Kroupa01}
and also simple power law distributions with different values for the slope $\alpha$ \citep[see][]{Salpeter55}.}
We fix the total number of stars to $N = 32768$,
the lower stellar mass limit to $m_{\rm low} = 0.2 \ \rm M_\odot$ and the upper mass limit to
$m_{\rm upper} = 150 \ \rm M_\odot$. Finally, we use values for the metallicity Z ranging from 0.0001 to 0.02.
We investigate in total 15 models with different slopes of the IMF and metallicities and for
each one of them we create a set of 100 random realisations. We evolve the
stars of each realisation to $10 \ \rm Myr$ using the stellar evolution code \texttt{sse} \citep{HPT00}
and we calculate {averages for the number of SBH created and also for their higher and lower masses.
The results are described in Table \ref{tab: table1}.}
Assuming no supernova kicks, the number of SBHs created
depends strongly on the choice of the IMF slopes and ranges from $\sim$ 20 to $\sim$ 70 in our models.
On the other hand the masses of the SBHs depend on the metallicity and range from $\sim 3 \ \rm M_\odot$
(for Z $= 0.02$) to $\sim 27 \ \rm M_\odot$ (for Z $< 0.001$).

\begin{table*}
\begin{center}
\caption{{Description of the full set of initial data created for the investigation of
the BH number and masses using different IMFs. We use a Kroupa '93 \citep{KroupaEtAl93}, a Kroupa '01 \citep{Kroupa00b},
a Salpeter \citep{Salpeter55} and two simple power law mass functions with slopes $\alpha=-2.5,-2.4$.
For each IMF we use three different values for the metallicity (Z), 0.0001, 0.001 and 0.02 and we create
100 realisations for each IMF-Z combination. We then evolve the stars up to 10 Myr using the stellar evolution code \texttt{sse}.
Finally, we find averages for the
number of BHs (third column) and their minimum (fourth column) and maximum (fifth column) masses.
In all data sets the total number of stars is 32768 and their initial masses range from $0.2 M_\odot $ to $150 M_\odot$.}}
\begin{tabular}{|c|c|c|c|c|}
 \cline{1-5}
{IMF} & {Z} & {$N_{\rm BHs}$} & {$M_{\rm BH \, max}$} & {$M_{\rm BH \, min}$} \\
 \cline{1-5}
Kroupa '93 & 0.02 & 26 $\pm$ 4& 14.11 $\pm$ 0.90 & 3.33 $\pm$ 0.28  \\ \cline{1-5}
Kroupa '93 & 0.001 & 22 $\pm$ 5& 25.58 $\pm$ 2.60 &  14.50 $\pm$ 0.22\\ \cline{1-5}
Kroupa '93 & 0.0001 & 23 $\pm$ 4 & 26.46 $\pm$ 0.64 & 15.35 $\pm$ 1.54 \\ \cline{1-5}

Kroupa '01 & 0.02 & 71 $\pm$ 9& 15.00 $\pm$ 0.26 &  3.14 $\pm$ 0.15\\ \cline{1-5}
Kroupa '01 & 0.001 & 68 $\pm$ 9 & 26.19 $\pm$ 0.16 & 13.88 $\pm$ 0.08  \\ \cline{1-5}
Kroupa '01 & 0.0001 & 69 $\pm$ 9& 26.88 $\pm$ 0.23& 14.13 $\pm$ 1.24  \\ \cline{1-5}

Salpeter & 0.02 & 54 $\pm$ 6  & 14.77 $\pm$ 0.47 & 3.19 $\pm$ 0.17  \\ \cline{1-5}
Salpeter & 0.001 & 50 $\pm$ 6 & 26.05 $\pm$ 0.30 & 13.38 $\pm$ 0.01 \\ \cline{1-5}
Salpeter & 0.0001 & 51 $\pm$ 6   &  26.74 $\pm$ 0.40   &  14.12 $\pm$ 1.51  \\ \cline{1-5}

Power Law ($\alpha=2.5$) & 0.02 & 29 $\pm$ 5 & 14.21 $\pm$ 0.85 & 3.38$\pm$ 0.32 \\ \cline{1-5}
Power Law ($\alpha=2.5$)& 0.001 & 26 $\pm$ 4 & 25.86 $\pm$ 0.52 & 14.88 $\pm$ 0.22 \\ \cline{1-5}
Power Law ($\alpha=2.5$)& 0.0001 & 27 $\pm$ 4 & 26.38 $\pm$ 0.62 & 15.66 $\pm$ 2.06  \\ \cline{1-5}

Power Law ($\alpha=2.4$) & 0.02 & 45 $\pm$ 6 &  14.60 $\pm$ 0.68 & 3.25 $\pm$ 0.24  \\ \cline{1-5}
Power Law ($\alpha=2.4$)& 0.001 & 41 $\pm$ 6  &   26.03 $\pm$ 0.28  &   13.11 $\pm$ 1.70 \\ \cline{1-5}
Power Law ($\alpha=2.4$)& 0.0001 & 42 $\pm$ 6  &   26.77 $\pm$ 0.32 &  14.47 $\pm$ 1.67  \\ \cline{1-5}
\cline{1-5}
\end{tabular}
\label{tab: table1}
\end{center}
\end{table*}

For the initial data of our simulations, we picked 4 representative cases from our investigation that
produce low and high numbers of SBHs. We also picked a
value Z = 0.001 for the metallicity as typical for a GC which resulted in the formation
of SBHs with masses between $\sim 13 \ \rm M_\odot$ and $27 \ \rm M_\odot$. In those models
all stars with masses above $20 \ \rm M_\odot$ have evolved off the main sequence at $ 10 \ \rm Myr$.

For our fiducial simulation A, we choose slopes $\alpha_1 = 1.3$ and $\alpha_2 = 2.4$,
which, after stellar evolution until $t = 10 \ \rm Myr$, result in $62$ stellar-mass
BHs in the system, close to the highest number of SBHs created in
our models. For the distribution of stars and BHs in the cluster, we use a King profile
\citep{King66} with concentration parameter $W_0 = 7$.
The initial escape velocity at the centre of the cluster is $\sim 17\,{\rm km\,s}^{-1}$.
At the centre of the cluster we introduce
an IMBH of mass $M_{\bullet} = 500\, \rm M_\odot$ and correct the velocities of all
stars and BHs of the GC to reach dynamical equilibrium.
We created also three additional initial data changing the IMF, the mass of the IMBH and/or the initial
concentration of the clusters. Case B is like A but with
$M_{\bullet} = 1,000\, \rm M_\odot$ and $\alpha_2 = 2.5$, which results in 52 SBHs;
case C is like B but with a King parameter of 6 and 48 SBHs. Finally, case D
is like A but with a King parameter of 6 and $\alpha_1 = 1.2$ and $\alpha_2 =
2.7$, which result in only 17 SBHs, close to the lower number of SBHs created in our test models.
{The initial data for the 4 simulations are described in Table \ref{tab: table2}.}
\begin{table*}
\begin{center}
\caption{{Initial data for the 4 simulations.}}
\begin{tabular}{|c|c|c|c|c|c|}
 \cline{1-6}
{Case} & $\alpha_{\rm 1}$ & $\alpha_{\rm 2}$ & $N_{\rm BHs}$ & $W{\rm_0}$ & $M_{\rm IMBH} \ [M_{\rm \odot}]$ \\
\cline{1-6}

A & 1.3 & 2.4 & 62 & 7 & 500   \\ \cline{1-6}
B & 1.3 & 2.5 & 52 & 7 & 1000   \\ \cline{1-6}
C & 1.3 & 2.5 & 48 & 6 & 1000   \\ \cline{1-6}
D & 1.2 & 2.7 & 17 & 6 & 500   \\ \cline{1-6}

\cline{1-6}
\end{tabular}
\label{tab: table2}
\end{center}
\end{table*}

We performed the dynamical evolution of each of the 4 models using \texttt{Myriad}, which
treats the stars and stellar remnants as point particles, but takes into account their sizes
 in the case of collisions. No primordial mass segregation is taken into
account, so the BHs are formed in all the distances from the centre of the cluster.
The absence of initial mass-segregation leads to an underestimate of the initial frequency of IMBH-SBH interactions,
but, as we will show below, most of the most massive SBHs sink to the center and interact with the IMBH very soon.
Stellar evolution is only used for creating the initial
data for our models. During the $N$-body simulations stellar evolution is turned off, so the masses of stars and
the masses and the number of remnants remain constant in time.
This is a simplification, which does dot have a significant impact on the dynamics
of the IMBH and therefore on our results. Further (i.e. after the $10 \ \rm Myr$ of the initial data)
stellar evolution would create a number of SBHs with very low mass ($< 10 \ \rm M_\odot$), which would have a
negligible influence on the dynamics of the IMBH and on a possible binary that the IMBH would form
with one of the higher-mass SBHs. Low-mass SBHs are expected not to be able to replace higher-mass SBH as
companions of the IMBH. Instead, they are expected to be ejected easily through natal kicks and interactions
with the IMBH and other higher-mass SBHs \citep{BaumgardtEtAl04b}.

From our set of simulations, only case A had an IMRI; we will therefore
focus on this case in the remainder of the article. As of now, \texttt{Myriad} runs only with the
assistance of the special-purpose GRAPE system \citep{GRAPE6}, so that we are subjected to the the
availability and performance of GRAPE systems.

\section{Dynamics of the system}
\label{sec:Dynamics}

\begin{figure}
\resizebox{\hsize}{!}
          {\includegraphics[scale=1,clip]{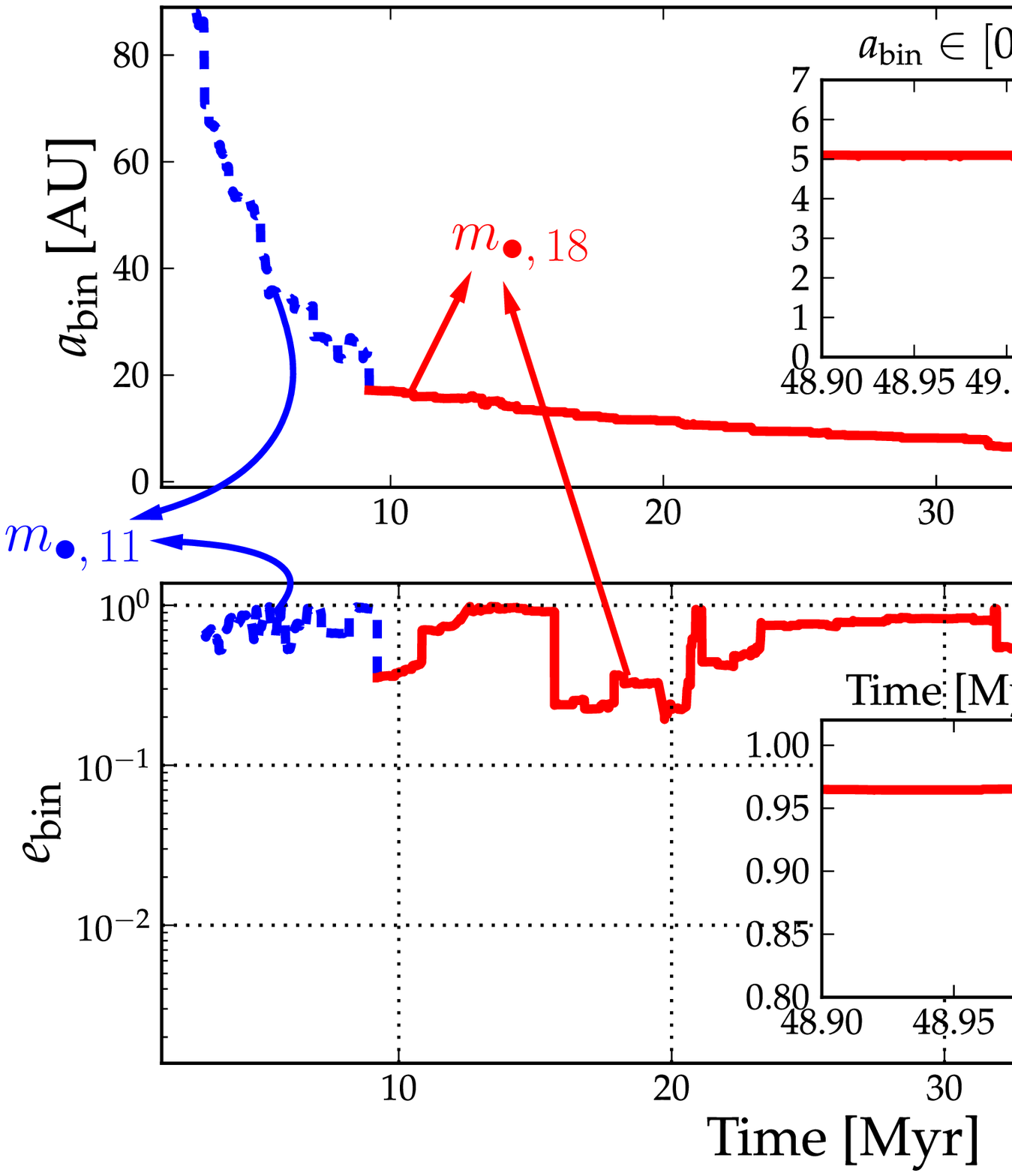}}
\caption
   {
Evolution of the semi-major axis and eccentricity of the different three
binaries formed with the IMBH.  Shortly after the beginning of the simulation,
the IMBH builds a binary with the SBH with the 11th most massive mass,
$m_{\bullet,\,11}$. This corresponds to the first part of the curve (dashed blue
curve). Later there is an interaction which leads to a companion exchange for
the binary, the SBH with the 18th most massive mass, $m_{\bullet,\,18}$. This
binary lives for about 40 Myr. We can see that the two first binaries have
phases of very high eccentricity, $e_{\rm bin} \sim 1$, but not high enough to
lead to a coalescence. The jumps in $e_{\rm bin}$ indicate that the IMBH-SBH is
still in a regime in which interactions with other stars play an important
role. The system shrinks further and further until there is a three-body
interaction. The binary is unbound and for a short period of time the IMBH has
no companion, as indicated in the zoom-in subplots embedded in both, the upper
and lower panels. Then the final binary forms, with the second most massive SBH.
This binary is very hard and quickly losses energy via GWs radiation, which
very efficiently leads to circularization and the final merger.
   }
\label{fig.Semi_eccentricity}

\end{figure}

Initially, the IMBH interacts strongly with a sub-group of stars and SBHs that
contains approximately $20$ members. As the system evolves, the members of
this sub-group change. Soon, most of the stellar-mass BHs of the system sink
towards the centre and start to interact with the IMBH and its environment.
During this process, some of them receive big kicks due to $3$-body
interactions and are slingshot away from the centre of the cluster or GC itself.
After $T \sim 3$ Myr the first stable IMBH-SBH binary forms. The companion of
the IMBH is a SBH with mass $m_{\bullet,\,11} = 23.9\,\rm M_{\odot}$ and the
initial semi-major axis of this binary is $a \sim 88$ AU. At $T \sim 9.2$ Myr
this binary has a strong interaction with another SBH of the system. The
interaction leads to a change of companion for the IMBH, which now builds a
binary with a SBH of mass $m_{\bullet,\,18} = 20.1\,\rm M_{\odot}$. The initial
semi-major axis of the new binary is $a \sim 17.6$ AU. This binary survives for
nearly $40$ Myr, but its characteristics vary significantly. At $T \sim 49$
Myr the semi-major axis changes to $a \sim 5$ AU, while the eccentricity
increases to $e=0.965$.  At this point in the simulation, this binary interacts
strongly with the second most massive SBH, which leads to a companion exchange.
The new binary has an initial semi-major axis of $a \sim 6.55$ AU and a very
high eccentricity, of $e=0.999$. The mass of the new companion SBH is
$m_{\bullet,\,2} = 26.54\,\rm M_{\odot}$. In Fig.(\ref{fig.Semi_eccentricity}) we
show the evolution of the semi-major axis and eccentricity for all of these
binaries combined into a single curve. After some $T \sim 13,000$ yr the binary
merges and the resulting IMBH receives a random recoil velocity that depends on
the mass ratio of the two members of the system and on the random spins that
the code assigned to them. This ``gravitational rocket'' or recoil is such that the resulting
velocity exceeds the escape velocity and the merged system leaves the GC.
This is due to the fact that we are using a low number of stars for the
clusters; more realistic clusters will have larger escape velocities, so that
the retained fraction of recoiling IMBH is larger and not well-represented by
our case.  We studied the distribution of recoil velocities for a merger of a
binary similar to that of simulation A.  We ran a two-body interaction $10^7$ times and
calculated the recoil using equation (\ref{recoil}) with different spin
orientations and magnitudes for the two black holes. We found that the most probable
recoil velocity for a binary such as the one of case A peaks around 25 ${\rm
km\,s}^{-1}$, with a probability of 21\% that the merged system achieves
velocities greater than $50 \ {\rm km\,s}^{-1}$, of the order of realistic GC escape
velocities.

\begin{figure}
\resizebox{\hsize}{!}
          {\includegraphics[scale=1,clip]{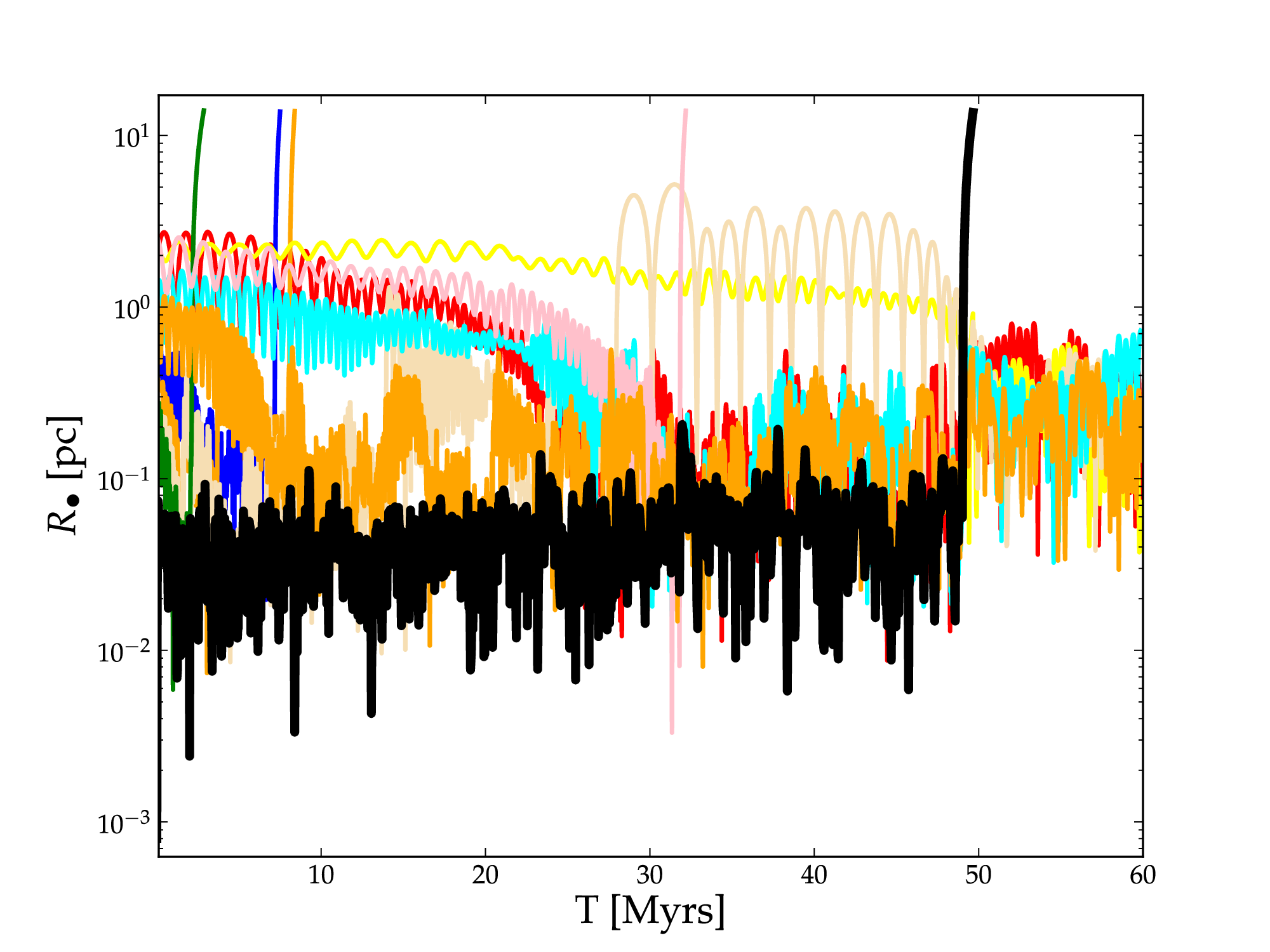}}
\caption
   {
Distance $R_{\bullet}$ to the density center of the GC of the ten most massive
SBHs and the IMBH (solid black line). Strong interactions of the SBHs lead to
ejection of four of them before the IMBH merges. They are removed from the
simulation when $R_{\bullet} > 10$ pc and they are unbound with the GC.  At $T
\sim 47.7$ Myr the IMRI leads to a coalescence that kicks the resulting merged
system off the GC.
   }
\label{fig.bh_positions_relative_to_center}
\end{figure}

In Fig. (\ref{fig.bh_positions_relative_to_center}) we show the evolution of
the distances of the $10$ most massive SBHs from the center. The SBHs inspiral
the center very rapidly, as long as the IMBH exists in the cluster. Some of
them escape the system, after passing very close to the central binary. After
the IMBH merges with its binary companion SBH, the coalesced system leaves the
GC and the trajectories of the remaining SBHs are not as steep, because they
orbit the center of density of the GC without sinking rapidly into it.

\begin{figure}
\resizebox{\hsize}{!}
          {\includegraphics[scale=1,clip]{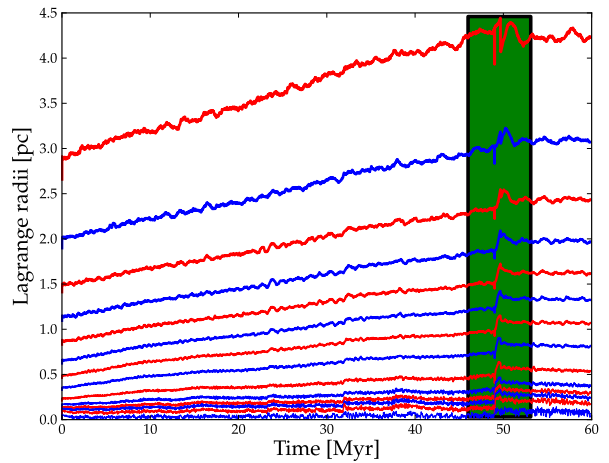}}
\caption
   {
Lagrangian radii showing the evolution of different mass fractions in the
cluster: from the top to the bottom 90, 80 ... 20, 10, 5, 3, 2, 1 and 0.1\% of the total
mass. The green rectangle shows the interval of time before and after the kick
of the IMBH off the cluster. All mass curves suffer a jump at the moment of
ejection. After the removal of the heating source from the center, the curves
are flatter and their slopes start to decrease.
   }
\label{fig.lagrange}
\end{figure}

In Fig. (\ref{fig.lagrange}) we show the Lagrange radii of the cluster during
the simulation. We stop the simulation at $\sim 10 \ \rm Myr$ after the ejection
of the IMBH. From t=0 until the ejection of the IMBH, which happened at $t \sim 49 \ \rm Myr$,
the Lagrange radii increase
constantly. This agrees with other results of other $N$-body \citep{BaumgardtEtAl04b}
and recent Monte Carlo \citep{UmbreitEtAl2012} simulations of GCs containing IMBHs, and it
happens because the central IMBH and the IMBH-SBH binary that
forms almost instantly after the beginning of the simulation, act as a heating
source for the cluster. Kinetic energy is transferred from the IMBH-SBH
binary to the stars and SBHs that pass close from the density center making
the binary constantly harder and the stars more energetic and thus expanding
the cluster. When the IMBH is removed from the cluster, the heating source
is absent, so the cluster starts contracting slowly as is obviously shown in
the Lagrange radii. The shrinkage of the cluster would continue until
the central number density of stars becomes high enough that
another heating source (i.e. a new binary) is formed at the center.
Even though we integrated the system for another $\sim 10 \ \rm Myr$ after
the ejection, no such source formed.

In all simulations the IMBH had as a companion a SBH, which was replaced
several times with another SBH usually of greater mass. In the end, the companion
of the IMBH was one of the most massive SBHs of the cluster. Most of the lower-mass
SBHs after sinking to the center and interacting with the central IMBH-SBH binary,
were kicked out of the GC. Also, in none of the simulations we find a
main-sequence or giant star tidally disrupted by the IMBH. This is in agreement
with the models of \citep{BaumgardtEtAl04b} which contain a number of massive SBHs. In our
models, apart from the IMBH, SBHs are the most massive objects in the GC and therefore
sink to the center faster than any other star. As a result a IMBH-SBH binary forms very soon
with the companion of the IMBH being more massive than any other non-SBH object
of the cluster. Thus, only interactions of the binary with another SBH of comparable or higher mass
than the current companion of the IMBH may lead to a companion exchange, so it
is almost impossible for a normal star to come too close to the IMBH to
get tidally disrupted. Therefore, tidal disruptions of stars are not favoured in our models.

After the merger, the IMBH leaves
the GC without any companion. This may be an artifact of the low number
of stars used in the simulation. In real clusters a small number of
stars or remnants are expected to be bound to the IMBH, so if the IMBH receives a large kick,
they will follow it outside the GC. In this case, and if there are no massive SBHs among the IMBH companions,
 some stars might be tidally disrupted by the IMBH, and thus the system might become
 a ULX not associated with a GC.

\section{Gravitational waves from an IMRI}
\label{sec:GW}

\begin{figure}
\resizebox{\hsize}{!}
          {\includegraphics[scale=1,clip]{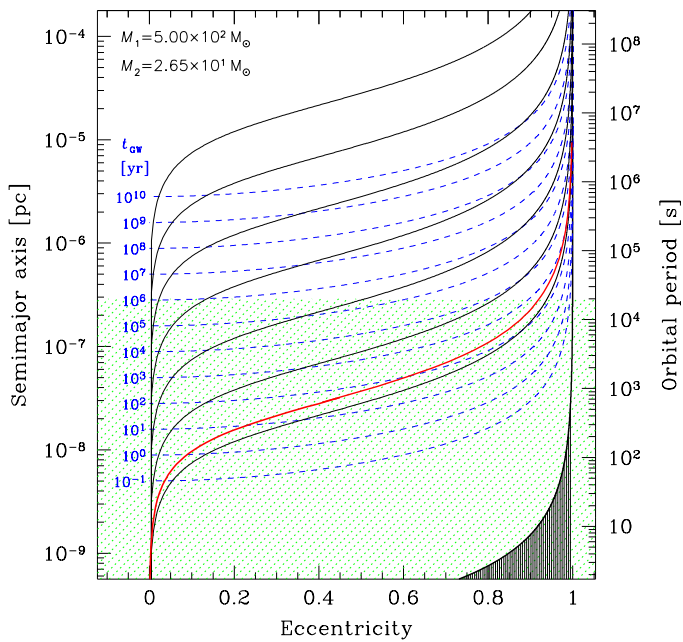}}
\caption
   {
Inspiral of the SBH into the IMBH from the top to the bottom and from the left
to the right in the eccentricity--semi-major axis plane.  The red, solid curve
starting at a very eccentric orbit shows the results of the {\sc Nbody}
simulation. The dashed, green region corresponds to the band of a LISA-like
mission.  Dashed, blue curves correspond to the trajectories due only to the
emission of GWs in the \cite{PM63} approximation. We also plot the
corresponding merger timescales in the same approximation in dashed, blue lines
starting at $10^{10}$ years, and in solid, black lines the corresponding
trajectories for evolution by GW emission \cite{Peters64} approximation. The
black-shaded region on the right corresponds to the last stable circular orbit.
Since the binary starts at a very high eccentricity, it basically follows one
of the solid black lines, because it merges quickly and does not interact with
other stars in the system.
   }
\label{fig.Evolution_Binary_Peters}
\end{figure}

In this section we follow the binary IMBH-SBH from the standpoint of emission
of GWs. In Fig.(\ref{fig.Evolution_Binary_Peters}) we can see the evolution of
the IMRI in a semi-major axis and orbital period -- eccentricity plane. The
binary enters the plot from the top with a high eccentricity, which places it
very close to the innermost stable circular orbit, but the loss of energy
quickly circularises it and drives it to lower eccentricities. As we discussed
in the previous section, the binary forms with a very small initial semi-major
axis, so that it hardens very efficiently. Hence, the binary follows closely
what we can expect from the approach of \cite{Peters64}, since the
post-Newtonian terms lead the evolution of the system, which can be regarded as
dynamically decoupled from the GC. It then enters the band of a LISA-like or
ALIA detector with a significant eccentricity and the simulation is
stopped when the semi-major axis is $a = 5\, \rm R_{\rm Schw}$, the Schwarzschild
radius of the IMBH. That is the moment at which the code
assigns a recoil velocity to the merged system based on the spins of the two
compact objects.

\begin{figure}
\resizebox{\hsize}{!}
          {\includegraphics[scale=1,clip]{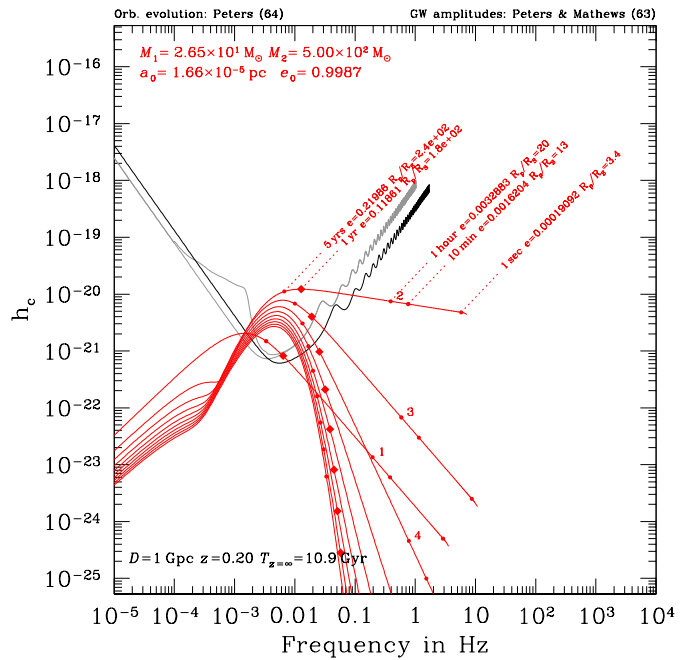}}
\caption
   {
Characteristic amplitude $h_c$ of the first harmonics of the quadruple
gravitational radiation emitted during the inspiral of the IMRI.
The numbers show the first four of the harmonics. The orbital
evolution is calculated in the \cite{Peters64} approximation and the amplitudes
as in \cite{PM63}.  We assume the source is at a distance $\rm D = 1\, \rm Gpc$. We
indicate with a solid curve the noise curve $\sqrt{f\,S_h(f)}$ for the ALIA
detector with an armlength of $3\times 10^9$ m, a telescope diameter of 0.58 m,
and a 1-way position noise of $8/\sqrt{\rm Hz}$ pm; i.e. the 3H configuration
of \cite{GongEtAl11}. We also add the noise curve for a LISA-like detector
\citep[in grey,][]{LHH00}, with the Galactic binary white dwarf confusion
background \citep{BH97}. Note that the SNR is not given by the height above the
curve, but by the area below it. For each panel we show the ratio $R_{\rm
p}^0/R_{\rm s}$, the initial periapsis distance over the Schwarzschild radius
of the system.  We indicate the moments in the evolution for which the time to
coalescence is 5, 1 yr, 1 hour, 10 minutes and 1 sec.
   }
\label{fig.Evolution_Binary_Peters_ALIA_LISA}
\end{figure}

In Fig.(\ref{fig.Evolution_Binary_Peters_ALIA_LISA}) we can see the same from the
perspective of the characteristic amplitude and frequency of the waves. We display the
first harmonics in the approximation of Keplerian ellipses of \cite{PM63}.

\section{Conclusions}
\label{sec:Conclusions}

In this work we have investigated with a direct-summation code the evolution of GCs that
harbour an IMBH in their center. The code uses relativistic corrections and a
prescription for gravitational recoil. For one of the cases we find that an
IMRI forms with a SBH due to close interactions, which leads to the
ejection of IMBH after coalescence. We follow the properties of the IMRI
from a standpoint of the global evolution of the cluster and of GW.

Before the formation of the IMRI and the subsequent ejection of the IMBH,
the cluster experiences strong expansion as a result of two-body
relaxation in the presence of an IMBH. The IMBH-SBH binaries that are
formed transfer kinetic energy to the stars that sink to the center and as a
result the GC expands significantly. Some of the inner Lagrange radii of
the cluster almost double in size during the first $\sim 50 \ \rm Myr$ of
dynamical evolution. Also, most of the massive SBHs sink to the center
very rapidly and after interacting with the IMBH, if they do not become
its companions, they receive large kicks and get ejected violently from the GC.
After the ejection of the IMBH, the GC slowly starts to contract as
a result of the absence of the heating source at the center. This
might lead to another core collapse of the GC, something we do not
observe in the simulation in the first $10 \ \rm Myr$ after the ejection
if the IMBH.

In our simulations, the IMBH forms a binary with a SBH very early and
then exchanges its companion several times. In the simulation we observed
an IMRI, the IMBH formed a binary with the second most massive SBH of the system.
The initial high eccentricity of the IMBH-SBH binary lead to an IMRI and a subsequent merger.
We showed that for $z \leqslant 0.7$ the energy loss of the binary in GWs is easily detectable by space-borne
missions such as a LISA-like observatory \citep{Amaro-SeoaneEtAl2012} or ALIA in its 8 pc
configuration. Moreover, the IMRI enters the bandwidth of the detectors with a
very high eccentricity, $e = 0.9987$, as with the EMRIs. One year before the
final coalescence, the system still retains a residual eccentricity of $e \sim
0.12$, and ten minutes before merger of $e \sim 10^{-3}$, which is detectable by
data-analysis techniques \citep{AS10a,PorterSesana10,KeyCornish11}.

IMRIs represent a test of GR, as well as a probe of space and time around
massive black holes and also of the innermost kinematics of GCs to very large
distances, of the order of a few Gpc. On the top of that, a successful
detection would represent very robust proof for the existence of IMBHs. The
fact that the kick is making the merged system leave the GC is possibly an
artifact of the low particle number we used in the simulations, though in
principle recoiling velocities can be much higher than the escape velocity of a
cluster, of the order of $\sim 50 \ {\rm km\,s}^{-1}$ \citep[see
e.g.][]{Rezzolla09,Holley-Bockelmann08}.
However, we have also demonstrated that there is a
non-negligible statistical probability that a similar case leads to a kick of
the IMBH off a realistic GC.

In our simulations we do not observe any tidal disruption of stars and also in the simulation
in which we had an IMRI, the IMBH left the system without any companion. This is probably also
an artifact of the low number of stars used in the simulations. In real clusters
normally there is a small number of stars bound to the IMBH, so after all the SBHs
are ejected and probably after some IMBH-SBH mergers, the IMBH will be
followed by a number of main-sequence or giant stars, even if it is ejected
from the GC. Therefore, if after the IMBH-SBH merger the IMBH remains in the GC, it
might soon become an X-ray source at the center of the cluster, similar to
the ULXs that are located at the centers of GCs \citep[e.g. the ULX at NGC 404][]{NylandEtAl2012,BinderEtAl2011}.
On the other hand, if
it escapes the GC, followed by the stars bound to it, it might soon become a ULX
outside of the GC. In this context, HLX-1 might be an escaping IMBH
that originated at the center of ESO243-49 and received a large recoil velocity after
a merger with a massive SBH. The kick is responsible for the escape if the
IMBH, which is followed by a number of stars gravitationally bound to it.
This scenario supports one of the possible scenarios for HLX-1 suggested
by \cite{SoriaEtAl2012} according to which the HLX-1 is an IMBH embedded
in a young population of stars with ages $< 10 \ \rm Myr$ and total mass with upper bound of
$\sim 10^4 \ \rm M_\odot$. A larger number of simulations of escaping IMBHs using
a realistic number of stars would be appropriate for testing this scenario.

In spite of the code been ported to run on a PC
with special-purpose hardware GRAPE, we can not cover a broader parameter
space, nor study cases with a larger number of stars, or study the global
dynamical evolution of the GC after the kick for a longer time. We plan on
performing a better parameter space exploration thanks to the availability of GPUs, which will allow us to address
the limitations we described above. This will allow us to investigate the
potential global structure of the GC after the kick, since the impact on the
cluster could in principle be a signature for the process. Also, it will
allow us to study also the properties and detectability of the escaping IMBHs
and their possible companions.

\begin{acknowledgements}

We thank Emma Robinson for comments on the manuscript and Nikolaos Stergioulas
for suggesting us to examine also the possibility of kicks.  SK and PAS are
indebted to Xuefei Gong, Shan Bai and Yun Kau Lau for conversations, their
hospitality in Beijing and for sharing with us the data for the sensitivity
curve of the ALIA detector.  PAS thanks the organizers of the 2011 summer
workshop at the Aspen Center for Physics for inviting him to attend.  SK and
PAS are thankful to Rainer Spurzem and Fukun Liu for the invitation to the
meeting in Lijiang on black holes and to visit the KIAA and NAOC in Beijing in
summer of 2011, where this work was finished.  The work of SK was funded by the
Deutsches Zentrum f{\"u}r Luft- und Raumfahrt (DLR; through the LISA Germany
project), also by the German Science Foundation (DFG) via SFB/TR7 on
``Gravitational Waves'' and by the FONDECYT Postdoctoral Fellowship number 3130570.  SK
would like to thank Kleomenis Tsiganis for discussions.
\end{acknowledgements}

\nocite{BHMMcMPZ03b}

\label{lastpage}
\end{document}